\documentclass{PoS}
\usepackage{epsfig}
\title{Lattice study of monopoles in the Electroweak theory}

\ShortTitle{Lattice study of monopoles in the Electroweak theory}

\author{B.L.G.Bakker\\
        Department of Physics and Astronomy, Vrije Universiteit,
Amsterdam, The Netherlands\\
        E-mail: \email{blg.bakker@few.vu.nl}}

\author{M.A.Zubkov\\
        ITEP, B.Cheremushkinskaya 25, Moscow, 117259, Russia\\
        E-mail: \email{zubkov@itep.ru}}

\author{\speaker{A.I.Veselov} \thanks{poster presented by A.I.Veselov}\\
        ITEP, B.Cheremushkinskaya 25, Moscow, 117259, Russia\\
        E-mail: \email{veselov@itep.ru}}

\abstract{We investigated numerically properties of Nambu monopoles in lattice
Electroweak theory at realistic values of $\alpha$ and $\theta_W$. Our choice
of parameters of lattice Lagrangian corresponds to large values of the Higgs
boson mass $M_H > 2 M_W$.
 We find that the density of Nambu monopoles cannot be predicted by the
choice of the initial parameters of Electroweak theory and should be considered
as the new external parameter of the theory. We also investigate the difference
between the versions of Electroweak theory with the gauge groups $SU(2)\otimes
U(1)$ and $SU(2)\otimes U(1)/Z_2$. We do not detect any difference at $\alpha
\sim \frac{1}{128}$. However, such a difference appears in the strong coupling
region and is related to the properties of monopoles constructed of the
hypercharge field. }

\FullConference{The XXV International Symposium on Lattice Field Theory\\
         July 30-4 August 2007\\
         Regensburg, Germany}

\begin{document}

\section{Introduction}

The qualitative lattice investigation of the properties of Nambu
monopoles\cite{Nambu} in the Standard Model has been performed both at zero and
finite temperatures in the unphysical region of large coupling constants in
\cite{BVZ}. Nambu monopoles are found to be condensed in the symmetric phase of
lattice theory (and above the Electroweak transition in the finite temperature
theory). Here we continue this investigation for realistic values of the
renormalized coupling constants ($\alpha \sim 1/128$ and $\theta_W = \pi/6$)
within the zero temperature theory.

Earlier we considered the appearance of an additional discrete symmetry in the
fermion sector of the Standard Model\cite{BVZ,Z2007,Z2007_3}. This additional
symmetry allows to define Standard Model with the gauge group $SU(3)\times
SU(2) \times U(1)/{\cal Z}$, where ${\cal Z}$ is equal to $Z_6$, or to one of
its subgroups: $Z_3$ or $Z_2$.
 The emergence of $Z_6$ symmetry in technicolor models was considered in
\cite{Z2007_2}.

Here we use two lattice realizations of the Electroweak theory: with the gauge
groups $SU(2) \times U(1)/Z_2$, and $SU(2) \times U(1)$, respectively.

\section{Lattice models under investigation}
We consider lattice Weinberg - Salam Model in quenched approximation. The model
contains the gauge field ${\cal U} = (U, \theta)$, where $ \quad U
 \in SU(2), \quad e^{i\theta} \in U(1)$ are
realized as link variables. The potential for the scalar field is considered in
its simplest form \cite{BVZ} in the London limit, i.e., in the limit of
infinite bare Higgs mass. From the very beginning we fix the unitary gauge.

For the case of $SU(2) \times U(1)/Z_2$ symmetric model we chose the action of
the form
\begin{eqnarray}
 S_g & = & \beta \!\! \sum_{\rm plaquettes}\!\!
 ((1-\mbox{${\small \frac{1}{2}}$} \, {\rm Tr}\, U_p \cos \theta_p)
 + \mbox{${\small \frac{1}{2}}$} (1-\cos 2\theta_p))+\nonumber\\
 && + \gamma \sum_{xy}(1 - Re(U^{11}_{xy} e^{i\theta_{xy}})),
\end{eqnarray}
where the plaquette variables are defined as $U_p = U_{xy} U_{yz} U_{wz}^*
U_{xw}^*$, and $\theta_p = \theta_{xy} + \theta_{yz} - \theta_{wz} -
\theta_{xw}$ for the plaquette composed of the vertices $x,y,z,w$.

For the case of the conventional $SU(2) \times U(1)$ symmetric model we use the
action
\begin{eqnarray}
 S_g & = & \beta \!\! \sum_{\rm plaquettes}\!\!
 ((1-\mbox{${\small \frac{1}{2}}$} \, {\rm Tr}\, U_p )
 + 3 (1-\cos \theta_p))+\nonumber\\
 && + \gamma \sum_{xy}(1 - Re(U^{11}_{xy} e^{i\theta_{xy}})).
\end{eqnarray}
In both cases the bare Weinberg angle is $\theta_W = \pi/6$, which is close to
its experimental value. The renormalized Weinberg angle is to be calculated
through the ratio of the lattice masses:  ${\rm cos} \, \theta_W = M_W/M_Z$.
The bare electromagnetic coupling constant $\alpha$ is expressed through
$\beta$ as $\alpha = 1/4 \pi \beta$. However, the renormalized coupling
extracted from the potential for infinitely heavy fermions differs from this
simple expression, as will be shown in the next sections.

\section{The results}

\subsection{Phase diagram}

\begin{figure}
\begin{center}
 \epsfig{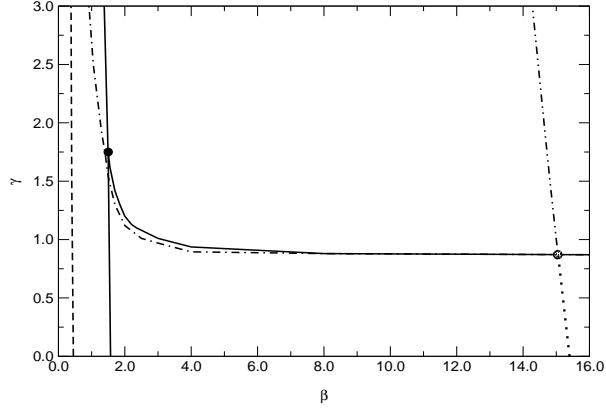}
\caption{\label{fig.1} The phase diagrams of the models in the
 $(\beta, \gamma)$-plane.}
\end{center}
\end{figure}

The phase diagrams of the two models under consideration are presented in
figure $1$.  The dashed vertical line represents the phase transition in the
$SU(2)\otimes U(1)$-symmetric model. This is the confinement-deconfinement
phase transition corresponding to the $U(1)$ constituents of the model. The
same transition for the $SU(2)\otimes U(1)/Z_2$-symmetric model  is represented
by the solid vertical line. The dashed horizontal line corresponds to the
transition between the broken and symmetric phases of model A. The continuous
horizontal line represents the same transition in model B. Interestingly, in
the $SU(2)\otimes U(1)/Z_2$ model both transition lines meet, forming a triple
point. Much attention was paid to this fact in \cite{BVZ}.

Real physics is commonly believed to be achieved within the phases of the two
models situated in the right upper corner of Fig.~$1$. The double-dotted-dashed
vertical line on the right-hand side of the diagram represents the line, where
the renormalized $\alpha$ is constant and equal to $1/128$.

All simulations were performed on lattices of sizes $8^4$ and $16^4$. Several
points were checked using a lattice $24^4$. In general we found no significant
difference between the mentioned lattice sizes.

\subsection{The masses}

The following variables are considered as creating a $Z$ boson and a $W$ boson,
respectively: $ Z_{xy}  =  Z^{\mu}_{x} \;
 = {\rm sin} \,[{\rm Arg} U_{xy}^{11} + \theta_{xy}]$, $ W_{xy}  =  W^{\mu}_{x} \,= \,U_{xy}^{12} e^{i\theta_{xy}}$.
In order to evaluate the masses of the $Z$-boson and Higgs boson we use the
zero - momentum correlators: $\sum_{\bar{x},\bar{y}}\langle \sum_{\mu}
Z^{\mu}_{x} Z^{\mu}_{y} \rangle   \sim
  e^{-M_{Z}|x_0-y_0|} + e^{-M_{Z}(L - |x_0-y_0|)}$, $\sum_{\bar{x},\bar{y}}\langle H_{x} H_{y}\rangle    \sim
  e^{-M_{H}|x_0-y_0|}+ e^{-M_{H}(L - |x_0-y_0|)} + const$.
In lattice calculations we used three different operators that create Higgs
bosons: $ H_x = \sum_{y} |W_{xy}|^2$, $H_x = \sum_{y} Z^2_{xy}$, and $H_x =
\sum_{y} Re(U^{11}_{xy} e^{i\theta_{xy}})$. In all cases $H_x$ is defined at
the site $x$, the sum $\sum_y$ is over its neighboring sites $y$.

After fixing the unitary gauge, lattice Electroweak theory becomes a lattice
$U(1)$ gauge theory. The $U(1)$ gauge field is $ A_{xy}  =  A^{\mu}_{x} \;
 = \,[-{\rm Arg} U_{xy}^{11} + \theta_{xy}]  \,{\rm mod} \,2\pi$.
The usual Electromagnetic field is $ A_{\rm EM}  =  A + Z^{\prime} - 2 \,{\rm
sin}^2\, \theta_W Z^{\prime}$,
where $Z^{\prime} = [ {\rm Arg} U_{xy}^{11} + \theta_{xy}]{\rm mod} 2\pi$.

The $W$ boson field is charged with respect to the $U(1)$ symmetry. Therefore
we fix the lattice Landau gauge in order to investigate the $W$ boson
propagator. The lattice Landau gauge is fixed via minimizing (with respect to
the $U(1)$ gauge transformations) of the following functional:
$ F  =  \sum_{xy}(1 - \cos(A_{xy})).$
Then we extract the mass of the $W$ boson from the correlator
$ \sum_{\bar{x},\bar{y}} \langle \sum_{\mu} W^{\mu}_{x} (W^{\mu}_{y})^{\dagger}
\rangle   \sim
  e^{-M_{W}|x_0-y_0|}+ e^{-M_{W}(L - |x_0-y_0|)}
$
The renormalized Weinberg angle is to be calculated through the ratio of the
lattice masses: ${\rm cos} \, \theta_W = M_W/M_Z$.

In the region $\beta \in (10,20)$, $\gamma \in (1,2)$ we found no difference
between the two versions of lattice Electroweak theory. Therefore, we omit
mentioning to what particular model the considered quantity belongs in this
region of coupling constants.

$W$-boson and $Z$-boson masses are found to change very slowly with the
variation of $\beta$.  The dependence on $\gamma$ seems to be stronger. Both
gauge boson masses grow with the decrease of $\gamma$.  We evaluate both masses
of $W$- and $Z$-bosons to be $0.23\pm 0.02$ at $\gamma = 1, \beta = 15$. We
cannot calculate the renormalized Weinberg angle at this point with reasonable
accuracy.

Unfortunately, the statistical errors do not allow us to calculate the Higgs
boson mass with a reasonable accuracy. Our data only allow us to draw the
conclusion that $M_H$ is larger than $2 M_W$.

\subsection{The renormalized coupling}

The bare constant $\alpha = e^2/4\pi$ (where $e$ is the electric charge) can be
easily calculated in our lattice model. It is found to be equal to $1/(4\pi
\beta)$. Therefore, its physical value $\alpha(M_Z)\sim 1/128$ could be
achieved at values of $\beta$ in some vicinity of $10$. This naive guess is,
however, to be corrected by the calculation of the renormalized coupling
constant $\alpha_R$. We perform this calculation using the potential for
infinitely heavy external fermions. We consider Wilson loops for the
right-handed external leptons: $
 {\cal W}^{\rm R}_{\rm lept}(l)  =
 \langle {\rm Re} \,\Pi_{(xy) \in l} e^{2i\theta_{xy}}\rangle.
$
Here $l$ denotes a closed contour on the lattice. We consider the following
quantity constructed from the rectangular Wilson loop of size $r\times t$:
$
 {\cal V}(r) = \lim_{t \rightarrow \infty}
 \frac{  {\cal W}(r\times t)}{{\cal W}(r\times (t+1))}.
$
At large enough distances we expect the appearance of the Coulomb interaction
$
 {\cal V}(r) = -\frac{\alpha_R}{r} + const.
$

The renormalized coupling constant $\alpha$ is found to be close to the
realistic value $\alpha(M_Z)=1/128$ along the line represented in Fig.~$1$.
Actually, a linear dependence of the potential for infinitely heavy
right-handed leptons on $1/r$ is observed already for $r=1$. Therefore we treat
this constant as $\alpha_R(1/a)=\alpha_R(100 {\rm GeV})\sim \alpha_R(M_Z)$.

\subsection{Nambu monopole density and percolation probability}

\begin{figure}
\begin{center}
 \epsfig{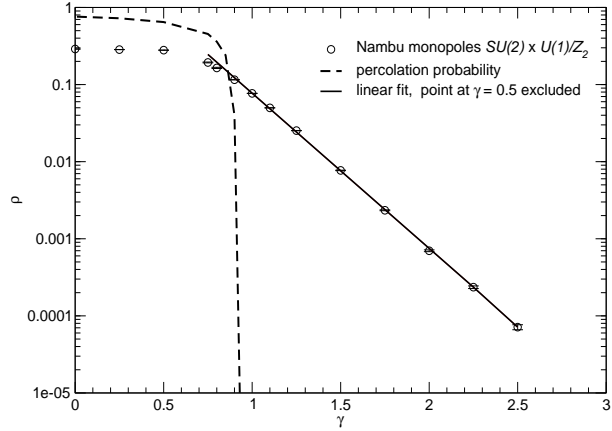}
\caption{\label{fig.5} Nambu monopole density and percolation probability as a
function of $\gamma$ along the line of constant $1/\alpha_R=128$.}
\end{center}
\end{figure}

According to \cite{Chernodub_Nambu} the worldlines of the quantum Nambu
monopoles could be extracted from the field configurations as follows:
\begin{equation}
 j_Z = \delta \Sigma = \frac{1}{2\pi} {}^*d([d Z^{\prime}]{\rm mod}2\pi)
\end{equation}
(The notations of differential forms on the lattice \cite{forms} are used
here.) The monopole density is defined as $
 \rho = \left\langle \frac{\sum_{\rm links}|j_{\rm link}|}{4L^4}
 \right\rangle,
$
where $L$ is the lattice size.

In order to investigate the condensation of  monopoles we use the percolation
probability $\Pi(A)$. It is the probability that two infinitely distant points
are connected by a monopole cluster (for more details of the definition see,
for example, \cite{BVZ1999}). In Fig.~$5$ we show Nambu monopole density and
percolation probability as a function of $\gamma$ along the line of constant
renormalized $\alpha_R = 1/128$. It is clear from Fig.~$5$ that the percolation
probability is the order parameter of the transition from the symmetric to the
broken phase. In order to compare the position of the transition between the
symmetric and broken phases with the point where percolation probability
vanishes,  we investigate the susceptibility $\chi = \langle H^2 \rangle -
\langle H\rangle^2$ extracted both from $H_Z = \sum_{y} Z^2_{xy}$ and $H_W =
\sum_{y} |W|^2_{xy}$ .

\begin{figure}
\begin{center}
 \epsfig{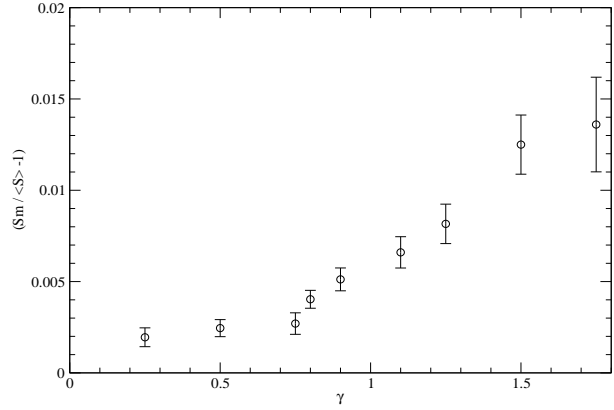}
\vspace{1.5ex} \caption{\label{fig.7} Plaquette action near monopole
trajectories along the line of constant $1/\alpha_R=128$.}
\end{center}
\end{figure}

The monopole worldline lives on the dual lattice. Each point of the worldline
is surrounded by a three - dimensional hypercube of the original lattice. We
measure the plaquette part of the action $S^{\rm mon}_{p}$ on the plaquettes
that belong to those $3$-dimensional hypercubes (normalized by the number of
such plaquettes). The excess of the plaquette action near monopole worldlines
over the mean plaquette part of the action $S_p$ is denoted by
$
\Delta S_p = \frac{1}{S_p}(S^{\rm mon}_{p} - S_p).
$
Very roughly $\Delta S_p$ can be considered as measuring the magnetic energy
(both $SU(2)$ and $U(1)$), which is carried by Nambu monopoles.

We also measure $S^{\rm mon}_{l}$, which is the part of the action $S^{\rm
mon}_{l}$ on the links of the original lattice that connect vertices of the two
incident $3$-dimensional hypercubes mentioned above. The excess of this link
action near monopole worldlines over the mean link part of the action $S_l$ is
denoted by
$
\Delta S_l = \frac{1}{S_l}(S^{\rm mon}_{l} - S_l).
$
For the simplicity of the calculations we get only one of the $8$ links that
connect incident hypercubes. The magnetic energy $\Delta S_p$ carried by Nambu
monopole is presented in Fig.~$7$. The behavior of both $\Delta S_p$ and
$\Delta S_l$ shows that a quantum Nambu monopole may indeed be considered as a
physical object.

\subsection{Evaluation of the lattice spacing}

The physical scale is given in our lattice theory by the value of the $W$-boson
mass $M^{phys}_W \sim 80$ GeV. Therefore the lattice spacing is evaluated to be
$a \sim [80 {\rm GeV}]^{-1} M_W$, where $M_W$ is the $W$ boson mass in lattice
units.

The real continuum physics should be approached along the the line of constant
$\alpha_R = \frac{1}{128}$, i.e. along the line of constant physics (at this
point we omit consideration of $\theta_W$ as according to our estimates it does
not vary crucially along this line). We investigated the dependence of the
ultraviolet cutoff $\Lambda = a^{-1} = (80~{\rm GeV})/M_W$ on $\gamma$ along
the line of constant physics. It occurs that $\Lambda$ is increasing slowly
along this line with decreasing $\gamma$ and achieves the value $350\pm 30$ GeV
at the transition point between the physical Higgs phase and the symmetric
phase. According to our results this value does not depend on the lattice size.
This means that the largest achievable value of the ultraviolet cutoff is equal
to $350 \pm 30$ GeV if the potential for the Higgs field is considered in the
London limit.

Our lattice study also demonstrates another peculiar feature of Electroweak
theory. If we are moving along the line of constant $\alpha=1/128$, then the
Nambu-monopole density decreases with increasing $\gamma$ (for $\gamma > 1$).
Its behavior is approximated with a nice accuracy by the simple formula: $\rho
\sim e^{2.08 - 4.6 \gamma}.$

Naively one may think that the density should decrease with increasing
ultraviolet cutoff. However, it occurs that the situation is inverse. This
means that the density of Nambu-monopoles is not fixed by the initial values of
the coupling constants and should be considered as {\it an additional
parameter} of Electroweak theory.

\section{Conclusions}

We investigated lattice Electroweak theory numerically at realistic values of
the coupling constants and for Higgs mass larger than $2 M_W$. We found that
the two definitions of the theory (with the gauge groups $SU(2)\otimes
U(1)/Z_2$ and $SU(2)\otimes U(1)$, rescpectively) do not lead to different
predictions at these values of the couplings.

Our investigation of the line of constant physics for the infinite bare self
coupling of the Higgs field allows us to draw the conclusion that the values of
lattice spacings smaller than $(350 \pm 30 {\rm GeV})^{-1}$ cannot be achieved
in principle for this choice of the potential for the Higgs field.

The action density near the Nambu monopole worldlines is found to exceed the
density averaged over the lattice in the physical region of the phase diagram.
This shows that Nambu monopoles can indeed be considered as physical objects.
Their percolation probability is found to be an order parameter for the
transition between the symmetric and broken phases. According to our numerical
data the density of Nambu monopoles in the continuum theory cannot be predicted
by the choice of the usual parameters of the Electroweak theory and should be
considered as a new external parameter of the theory.

\begin{acknowledgments}

This work was partly supported by RFBR grants 05-02-16306, and 07-02-00237,
RFBR-DFG grant 06-02-04010, by Federal Program of the Russian Ministry of
Industry, Science and Technology No 40.052.1.1.1112, by Grant for leading
scientific schools 843.2006.2.

\end{acknowledgments}

\end{document}